\documentstyle{article}

\newcommand{\Section}[1]{\section{#1}\setcounter{equation}{0}}

\def\QED{\hbox{\kern 1pt\vrule width 3pt height 7pt}}
\def\p{\partial}
\def\db{\bar\partial}

\def\d{\delta}
\def\dg{\dagger}

\def\b{\bar}

\def\be{\begin{equation}}
\def\bea{\begin{eqnarray}}

\def\nn{\nonumber}

\def\l{\label}
\def\ee{\end{equation}}
\def\eea{\end{eqnarray}}
\def\C{\rm {I\kern-.520em C}}

\def\o{\over}
\begin{document}
\begin{titlepage}
\hfill
\vbox{
    \halign{#\hfil         \cr
           hep-th/0109066 \cr
           } 
      }  
\vspace*{3mm}
\begin{center}
{\large \bf Axial Gauged Noncommutative U(2)/U(1) Wess-Zumino-Witten Model}
\footnote{
based on paper hep-th/0008120 with S. Parvizi}
\vskip .5in
{\bf Amir Masoud Ghezelbash}\footnote{
amasoud@theory.ipm.ac.ir}
\vskip .25in
{\em
Department of Physics, Az-zahra University,
Tehran 19834, Iran.}\\
\vskip .5in
\end{center}
\begin{abstract}
We construct various kinds of gauged
noncommutative WZW models. In particular, axial gauged noncommutative
$U(2)/U(1)$ WZW model is studied and by integrating
out the gauge fields, we obtain a noncommutative non-linear $\sigma$-model.
\end{abstract}
\end{titlepage}
\newpage
\def \dbar {\bar \partial}
\def \d  {\partial}
\Section{Introduction}
Noncommutative field theory has emerged from
string theory
in certain backgrounds \cite{CONNES,CHEUNG,SEIBERG,ARDALAN}. The
noncommutativity of space is defined by
the relation,
\be [x^\mu ,x^\nu ]=i\theta ^{\mu\nu},
\ee
where $\theta ^{\mu\nu}$ is a second rank antisymmetric real constant tensor.
The function algebra in the noncommutative space is defined by the
noncommutative and associative Moyal $\star $-product,
\be
(f\star g)(x)=e^{{i\over 2}\theta _{\mu\nu}{\partial\over {\partial \xi ^\mu}}
{\partial\over {\partial \eta ^\nu}}}f(\xi)g(\eta)|_{\xi =\eta =x}.
\ee

A noncommutative field theory is simply obtained by replacing
ordinary multiplication of functions by the Moyal $\star $-product.
An interesting field theory whose noncommutative version would be of
interest is the WZW model.
In \cite{DABROWSKI}, a noncommutative
non-linear $\sigma$-model has been studied and an infinite dimensional
symmetry
is found. They also derived some properties of noncommutative
WZW model. In \cite{FURUTA}, the $\beta$-function of the $U(N)$ noncommutative
WZW model was calculated and found to be the same as that of ordinary
commutative WZW model. Hence, the conformal symmetry in certain fixed points
is recovered. In \cite{MORENO} and \cite{CHU}, the
derivation of noncommutative WZW action from a gauge theory was carried out.
The connection between noncommutative two-dimensional fermion models and
noncommutative WZW models was studied in \cite{NEW,NEWW}.

In this letter, we study the gauged noncommutative 
WZW models. In section 2,
after a brief review of noncommutative WZW model,
we construct different versions of gauged noncommutative WZW models.
In section 3, we consider the axial gauged noncommutative $U(2)$ WZW model by
its diagonal
$U(1)$ subgroup. The obtained gauged action contains infinite derivatives
in its $\star$ structure and hence is a nonlocal field theory. Integration
over the gauge fields requires solving an integral equation which we solve
by perturbative expansion in $\theta$. The result is a
noncommutative non-linear $\sigma$-model, which may contains singular
structures or a black hole.

\Section{Different Versions of Gauged Noncommutative WZW Models}
The action of the noncommutative WZW model is \cite{MORENO}:
\be \l{action1}
S(g)={k\over 4\pi}\,\int
_{\Sigma}\,d^2z\,
Tr(\,g^{-1}\star \partial g\star g^{-1}\star  \bar \partial
g\,)-{k\over 12\pi}\,\int _{M}\,
Tr(\,g^{-1}\star dg\,)_\star ^3,
\ee
where $M$ is a three-dimensional manifold whose boundary is $\Sigma$,
and $g$ is a map from $\Sigma$ (or from its extension $M$) to the group $G$.
We assume that the coordinates $(z,\bar z)$ on the worldsheet $\Sigma$ are
noncommutative
but the extended coordinate $t$ on the manifold $M$  commutes with others:
\be
[z,\bar z]=\theta ,\quad [t,z]=[t,\bar z]=0.
\ee
We define the group-valued field $g$ by,
\be
g=e_\star ^{i\pi ^aT_a}= 1+i\pi ^aT_a+{1\over 2}(i\pi ^aT_a)_\star ^2+\cdots,
\ee
where the $T_a$'s are the generators of group $G$.

Inserting the $\star $-product of two group elements in the eq.
(\ref{action1}), we find the noncommutative Polyakov-Wiegmann identity,
\be\label{PW}
S(g\star h)= S(g) + S(h) + {1\over{16\pi}} \int d^2z Tr(g^{-1}\star
\bar\partial g \star
\partial h \star h^{-1}),
\ee
which is the same as ordinary commutative identity with
products replaced by $\star $-products.

Using the Polyakov-Wiegmann identity, we can show that the
action (\ref{action1}) is invariant under the following transformations:
\be\label{symm}
g(z,\bar z)\rightarrow h(z)\star  g(z,\bar z) \star  \bar h(\bar z).
\ee
The corresponding conserved currents
are, \bea
J(z) &=& {k\over{2\pi}}\bar\partial g\star  g^{-1},\nonumber\\
\bar J(\bar z) &=& {k\over{2\pi}}g^{-1} \star  \partial g.
\eea
By use of the equations of motion, we can show that these currents are
indeed conserved,
\be \l{EQMO}
\bar\p J(z)= \p \bar J(\b z)= 0.
\ee
The quantization of the noncommutative WZW model was done in \cite{PARVIZI},
and the current algebra of the noncommutative WZW model found to be,
\be
[J_a(\sigma),J_b(\sigma')]=\delta (\sigma -\sigma ')if_{ab}^c J_c(\sigma)
+
{k\over {2\pi}}i\delta' (\sigma -\sigma ')\delta_{ab},
\ee
and a similar relation for commutation of $\b J$'s.

Note that in the above commutation relation, $\theta$ does not appear,
and it is just as commutative ordinary affine algebra with the same
central charge. The absence of $\theta$ has been expected, since the
currents are holomorphic by equations of motion and hence commutative in
the sense of $\star $-product.

Constructing the energy momentum tensor is also straightforward,
\be\label{TENSOR}
T(z)= {1 \over {k+N}} :J_i(z)J_i(z):+{1\over k}:J_0(z)J_0(z):,
\ee
where $J_i$'s are $SU(N)$ currents and $J_0$ is the $U(1)$ current
corresponding to the subgroups of $U(N)=U(1)\times SU(N)$.
Again the  products in (\ref{TENSOR}) are commutative because of
holomorphicity
of the currents. So the Virasoro algebra is also the same as usual
standard form and its central charge is unchanged,
\be
c={{kN^2+N}\over{k+N}}.
\ee

We want to gauge the chiral symmetry (\ref{symm}) as,
\be \label{gtrans}
g(z,\bar z)\rightarrow h_L (z,\bar z)\star  g(z,\bar z) \star  h_R(z,\bar
z),
\ee
where $h_L$ and $h_R$ belong to $H$ some subgroup of $G$.
For finding the invariant action under the above transformation we need to add
gauge fields terms to the action (\ref{action1}) as follows:
\be
S(g,A,\b A)=S(g) + S_A +S_{\bar A} + S_2 + S_4,
\ee
where, $S(g)$ is the action (\ref{action1}) and
\bea\label{gaugeterm}
S_A &=& {k\over 4\pi}\,\int
\,d^2z\,Tr(\,A_L \star \bar\partial g\star g^{-1}\,), \nn\\
S_{\bar A} &=& {k\over 4\pi}\,\int
\,d^2z\,Tr(\,\bar A_R\star g^{-1} \star \partial g\,), \nn\\
S_2 &=& {k\over 4\pi}\,\int
\,d^2z\,Tr(\,\bar A_R\star A_L\,), \nn\\
 S_4 &=& {k\over 4\pi}\,\int
\,d^2z\,Tr(\,\bar A_R\star g^{-1} \star  A_L \star g\,).
\eea
Gauge transformations for the gauge fields are
\bea\label{atrans}
A_L &\rightarrow& h_L \star  (A_L + d) \star  h_L^{-1},\nn\\
A_R &\rightarrow& h_R^{-1} \star  (A_R + d) \star  h_R.
\eea

Using the Polyakov-Wiegmann identity (\ref{PW}), one can find the
ransformed form of $S(g)$ and the gauge field terms (\ref{gaugeterm}),
under the transformations (\ref{gtrans}) and (\ref{atrans}) \cite{PARVIZI}.

To find an invariant action $S(g,A,\b A)$, we have to choose
constraints on the subgroup elements $h_L$ and $h_R$. The first consistent
choice is,
$
h_R= h_L^{-1} \equiv h,
$,
and yields to following transformations,
\bea \l{GFV}
g &\rightarrow& g'=h^{-1}\star g\star h, \cr
A&\rightarrow&A'=h^{-1}\star (A\star h+\d h ),\cr
\b A&\rightarrow&\b A'=h^{-1}\star (\b A \star  h+\db h).
\eea
The corresponding invariant action, called vector gauged WZW action,
is,
\be \l{SVECTOR}
S_V(g,A,\b A)=S(g)+S_A-S_{\b A}+S_2-S_4.
\ee

The second choice is to take $h_L=h_R \equiv h$ with $h$ belonging to
an Abelian subgroup of $G$. In this case we find the following gauge
transformations,
\bea \l{GFA}
g &\rightarrow& g'=h\star g\star h, \cr
A&\rightarrow&A'=h\star (A\star h^{-1}+\d h^{-1}),\cr
\b A&\rightarrow&\b A'=h^{-1}\star (\b A \star  h -\db h),
\eea
with the so called axial gauged WZW action,
\be \l{SAXIAL}
S_A(g,A,\b A)=S(g)+S_A+S_{\b A}+S_2+S_4.
\ee

By integrating out the $A$ and $\b A$ from the actions (\ref{SVECTOR}) and
(\ref{SAXIAL}), in principle, we find the effective actions as
noncommutative non-linear $\sigma$-models.

\Section{Axial Gauged Noncommutative U(2)/U(1) WZW Model}

We take here the noncommutative gauged axial $U(2)$ WZW action
(\ref{SAXIAL}), gauged by the subgroup $U(1)$ diagonally embedded in
$U(2)$. The group element of $U(2)$ is
\be
g=\pmatrix{a_1&a_2\cr a_3&a_4},
\ee
with the following constraints,
\bea    \label{CONST}
a_1\star a_1^\dg+a_2\star a_2^\dg&=&1,\cr
a_3\star a_3^\dg+a_4\star a_4^\dg&=&1,\cr
a_1\star a_3^\dg+a_2\star a_4^\dg&=&0.
\eea
The gauge parts of the action (\ref{SAXIAL}) is
\be     \label{SGAUGE}
S_{gauge}={k \o {2\pi}} \int d^2z \{A\sum _i\b\p a_i\star a_i^\dg+
\sum _ia_i^\dg\star \p a_i \b A+2A\b A+\sum_i A\star a_i\star \b A\star
a_i^\dg \}. \ee
To illustrate the integrating over the gauge fields $A$ and $\b A$, we
consider abbreviated notations as
follows:
\bea\label{INTEG}
\int {\cal D}A{\cal D}\b A e^{-S_{gauge}}&=&
\int {\cal D}A{\cal D}\b A e^{-\int d^2z(A\star {\cal O}\star \b A+b\star
\b A+A\star \b b)}\nn\\
&=&\int {\cal D}A{\cal D}\b A e^{-\int d^2z\big( (A+b')\star {\cal O}\star (\b
A+\b b')-b'\star {\cal O}\star \b b'\big)}, \eea
where
\be \label{BB} 
b'\star {\cal O}=b,\qquad
{\cal O}\star \b b'=\b b.
\ee
The result of integration would be
\be       \label{EFFECT}
e^{-S_{eff}}=(\det {\cal O})^{-1/2}e^{\int d^2zb'\star \b b}e^{-S(g)}.
\ee
By comparing eq. (\ref{INTEG}) with (\ref{SGAUGE}), $b$ and $\b b$ could be read
as follows:
\be
b={{k}\over {2\pi}}\sum _i a_i^\dagger\star \p a_i,\qquad
\b b={{k}\over {2\pi}}\sum _i \b \p a_i\star a_i^\dagger ,
\ee
and ${\cal O}$ can be read from quadratic terms of gauge fields $A$ and $\b A$
in (\ref{SGAUGE}). In fact
by using the Fourier transformation of Moyal $\star $-products of functions,
the explicit Fourier transform of ${\cal O}$ is as follows:
\bea
{\cal O}(p_1,p_2)=2 e^{i(p_1\wedge p_2)} \delta (p_1+p_2)
&+&\int a_i(p_3)a_i^\dg (p_4) e^{i(p_1\wedge p_2+p_1\wedge p_3+p_1\wedge p_4
+p_2\wedge p_4+p_3\wedge p_4+p_3\wedge p_2)}\nn\\
& &\times \delta (p_1+p_2+p_3+p_4)
dp_3dp_4.
\eea
To find $b'$, we need to inverse the ${\cal O}$ operator in (\ref{BB}), and
this is equivalent to solve Fourier transform of (\ref{BB}) which is an
integral equation as follows:
\be \label{INTEQ}
2b'(p)+\int b'(p-p_1-p_2)a_i(p_1)a_i^\dg (p_2)
e^{-i(p\wedge p_2+p_1\wedge p+p_2\wedge p_1)}
dp_1dp_2=b(p),
\ee
where
\be
b(p)=i \int a_i^\dg (p-p_1)a_i(p_1)p_1
e^{i p\wedge p_1}dp_1.
\ee
To solve eq. (\ref{INTEQ}), we expand the $b'$ and exponential factors in
terms of $\theta$,
$
b'(p)=b'_0(p)+\theta b'_1(p)+\theta ^2 b'_2(p)+\cdots .
$ One finds \cite{PARVIZI},
\be \l{BSEFR}
b'_0(z, \b z)={1 \over 4}a_i ^\dagger \p a_i,\qquad
b'_1(z, \b z)=\p \p a_i \b \p a_i^\dg
-\b \p \p a_i \p a_i^\dg).
\ee
In obtaining the above expressions, we have used the unitarity
conditions
(\ref{CONST}) and the equations of motion (\ref{EQMO}).
The effective action arising from the $\int d^2z b'\star \b b$ term in
eq. (\ref{EFFECT}) could be found as a power series in
$\theta$,
\be
S_{eff}=S(g)+{1 \o 2}Tr \ln ({\cal O}) +S_{eff}^{(0)}+\theta
S_{eff}^{(1)}+\cdots , \ee
in which
\footnote{The term ${1\o 2}Tr \ln ({\cal O})$ gives the
noncommutative effective action for dilaton field.}
\bea \l{ACTIONS}
S_{eff}^{(0)}&=&-{{k}\over {8\pi}}\int d^2z a_i^\dagger \p a_i \b \p
a_j a_j^\dagger,\nn\\
S_{eff}^{(1)}&=&-{{k}\over {8\pi}}\int d^2z \big( (
\p a_i^\dagger \p  a_i+ a_i^\dg \p \p a_i)
(\b\p\b \p a_j  a_j^\dg+\b \p a_j \b \p a_j^\dg)\nn \\
&+& a_i^\dg\p a_i(\p \b\p a_j\b \p a_j^\dg-\b \p\b\p a_j\p a_j^\dg) 
-
(\b \p \p a_i \p a_i^\dg
-\p \p a_i \b \p a_i^\dg ) \b \p a_ka_k^\dg \big) .
\eea
By looking at equations (\ref{BSEFR}) and (\ref{ACTIONS}), we
suggest the following exact forms for $b'(z,\b z)$ and $S_{eff}$,
\bea \l{EXACT}
b'(z, \b z)&=&{1 \over 4}\p a_i \star a_i ^\dagger, \nn\\
S_{eff}&=&S(g)+{1 \o 2}Tr \ln ({\cal O})
-{{k}\over {8\pi}}\int d^2z \p a_i \star  a_i^\dagger \star \b \p
a_j \star a_j^\dagger .
\eea
We have to fix the gauge freedom by a gauge fixing condition
on $a_i$'s. Under infinitesimal axial gauge transformations (\ref{GFA}),
we find,
$
a_i'=a_i+a_i\star \epsilon+\epsilon \star a_i ,
$
in which $\epsilon$ is the infinitesimal parameter of the gauge transformation.
We can find ( at least perturbatively in $\theta$ ) some $\epsilon$ such that
${\Re (a_1')}=0$ and one may take this relation as the gauge fixing
condition. It is worth mentioning that the $U(2)/U(1)$ model after applying
all conditions gives us a three dimensional non-linear
noncommutative $\sigma$-model. The geometrical study of this target space
which may contains singular
structure will be interesting \cite{GH}.

\end{document}